\documentclass[conference,10pt]{IEEEtran}
\IEEEoverridecommandlockouts
\usepackage{cite}
\usepackage{amsmath,amssymb,amsfonts}
\usepackage{graphicx}
\usepackage{textcomp}
\usepackage{xcolor}
\usepackage{enumitem}
\usepackage{tikz}
\usepackage{algorithm}
\usepackage{algpseudocode}
\usepackage{array}
\usepackage{subcaption}
\usepackage{euscript}
\usepackage{multirow}
\usepackage{enumitem}
\algnewcommand\algorithmicforeach{\textbf{for each}}
\algdef{S}[FOR]{ForEach}[1]{\algorithmicforeach\ #1\ \algorithmicdo}

\usepackage[hyphens]{url}
\usepackage[pdfusetitle, pdfauthor={Michael Shell, My institution}]{hyperref}
\hypersetup{breaklinks=true}
\def\BibTeX{{\rm B\kern-.05em{\sc i\kern-.025em b}\kern-.08em
    T\kern-.1667em\lower.7ex\hbox{E}\kern-.125emX}}
\begin{document}

\title{\huge Secured Communication Schemes for UAVs in 5G: CRYSTALS-Kyber and IDS}
\author{\IEEEauthorblockN{Taneya Sharma\IEEEauthorrefmark{1}, Seyed Ahmad Soleymani\IEEEauthorrefmark{1}, Mohammad Shojafar\IEEEauthorrefmark{1}, and Rahim Tafazolli\IEEEauthorrefmark{1}}
\IEEEauthorblockA{\IEEEauthorrefmark{1} 5G/6GIC, Institute for Communication Systems (ICS), University of Surrey, Guildford, UK \\\{tt00851,s.soleymani,m.shojafar,r.tafazolli\}@surrey.ac.uk}
}
\maketitle

\begin{abstract}
This paper introduces a secure communication architecture for Unmanned Aerial Vehicles (UAVs) and ground stations in 5G networks, addressing critical challenges in network security. The proposed solution integrates the Advanced Encryption Standard (AES) with Elliptic Curve Cryptography (ECC) and CRYSTALS-Kyber for key encapsulation, offering a hybrid cryptographic approach. By incorporating CRYSTALS-Kyber, the framework mitigates vulnerabilities in ECC against quantum attacks, positioning it as a quantum-resistant alternative.

The architecture is based on a server-client model, with UAVs functioning as clients and the ground station acting as the server. The system was rigorously evaluated in both VPN and 5G environments. Experimental results confirm that CRYSTALS-Kyber delivers strong protection against quantum threats with minimal performance overhead, making it highly suitable for UAVs with resource constraints. Moreover, the proposed architecture integrates an Artificial Intelligence (AI)-based Intrusion Detection System (IDS) to further enhance security. In performance evaluations, the IDS demonstrated strong results across multiple models with XGBoost, particularly in more demanding scenarios, outperforming other models with an accuracy of 97.33\% and an AUC of 0.94. These findings underscore the potential of combining quantum-resistant encryption mechanisms with AI-driven IDS to create a robust, scalable, and secure communication framework for UAV networks, particularly within the high-performance requirements of 5G environments.

\end{abstract}


\section{Introduction}
In an era where UAVs play a pivotal role across various sectors, such as agriculture, disaster management, surveillance, and delivery services, ensuring secure communication between UAVs and ground stations has become paramount. The advent of 5G technology offers faster connections but also introduces new security challenges \cite{YAACOUB2020100218}, including vulnerabilities to cyber-attacks, increased risk of data interception, and the potential exploitation of network infrastructure weaknesses. To mitigate these security concerns, several approaches have been proposed. One approach involves using trust-based security schemes to filter out malicious UAVs and ensure data integrity within 5G networks \cite{9351702}. 

Despite the advancements in cryptographic and trust-based mechanisms, significant challenges persist in UAV communication networks. These include ensuring lightweight cryptographic implementations that do not overload resource-constrained UAV systems \cite{sabuwala2023}, addressing the latency introduced by encryption and IDS techniques, and enhancing the detection accuracy of AI-driven IDS without compromising real-time communication. Cecchinato et al.'s 2023 \cite{10129046} work on AES encryption for UAV communications demonstrates the usage of AES encryption. This paper extends the research by incorporating ECC and CRYSTALS-Kyber for Key Encapsulation Mechanism (KEM) alongside AES encryption to enhance security against quantum threats. In addition, the paper explores AI-based IDS to secure UAV communications in both VPN and 5G environments.

\noindent\textbf{Main Contributions:}
The primary focus of this paper can be summarized as follows:
\begin{enumerate}
    \item \textbf{Introduction of ECC and Post-Quantum Cryptography:} This paper introduces both ECC and CRYSTALS-Kyber for KEM with AES encryption, addressing quantum vulnerabilities and extending previous research that focused solely on AES encryption. 
    \item \textbf{Comprehensive Performance Evaluation:} A rigorous ECC and CRYSTALS-Kyber performance evaluation is conducted in VPN and 5G environments, offering insights into the trade-offs between quantum resistance and performance.
    \item \textbf{AI-Driven IDS Analysis:} Using simulated datasets, the research demonstrates the effectiveness of machine learning models such as Naive Bayes, Logistic Regression, and XGBoost in detecting intrusions with high accuracy.
    \item \textbf{Scalability and Real-World Application:} The proposed system architecture is lightweight and scalable, making it suitable for real-world applications, particularly in resource-constrained environments like UAVs integrated with Single Board Computers (SBCs).
\end{enumerate}

\noindent \textbf{Organization: }The rest of the paper is structured as follows. Section~\ref{sec:2} discusses related work on encryption methods and AI-based IDS, focusing on advancements in quantum-resistant cryptography and IDS techniques. Section~\ref{sec:3} presents the proposed system architecture, detailing the integration of AES encryption with ECC and CRYSTALS-Kyber for key encapsulation. Section~\ref{sec:4} provides a comprehensive performance evaluation of the encryption mechanisms and AI-based IDS models. Finally, Section~\ref{sec:5} concludes the paper and suggests potential directions for future research.

\section{Related Work}\label{sec:2}
This section reviews relevant work on the development of security frameworks for UAV communication, focusing on lightweight KEMs like ECC \cite{9365029} and how its limitations \cite{Shor_1997} against quantum attacks can be overcome using techniques like CRYSTALS-Kyber, which can also be implemented in a lightweight weight environment as presented in \cite{HE2024167, 8406610}. Additionally, the application of AI algorithms in IDS to enhance UAV security is discussed.

Rapid utilization of UAVs, particularly in military and defense sectors, has necessitated advancements in secure communication protocols. Cecchinato et al. (2023) presented a framework using AES encryption for real-time multimedia data transmission in UAVs. Their work demonstrated the effectiveness of lightweight AES encryption on UAV systems with limited computational resources, particularly in challenging radio conditions. However, the study did not address vulnerabilities to quantum attacks, which are becoming increasingly relevant with advancements in quantum computing \cite{10129046}.

Research has shifted towards quantum-resistant algorithms to address the shortcomings in ECC and other classical cryptographic systems. Alkim et al. introduced the CRYSTALS-Kyber algorithm, a post-quantum KEM, which has since gained attention for its resilience against quantum attacks and efficiency in cryptographic operations \cite{197151}. However, it does not demonstrate its effectiveness under a lightweight environment in the 5G communication channel.

In addition to encryption mechanisms, IDS has become crucial in securing UAV communication networks. Studies such as Wu et al. have highlighted the importance of lightweight IDS models, particularly for UAVs operating in resource-constrained environments like military and critical infrastructure \cite{10418859}. Their work introduces a tiny machine-learning model that addresses the computational limitations of UAVs while providing robust intrusion detection. Similarly, research on the effectiveness of XGBoost for IDS has shown promising results in accuracy and efficiency, making it a strong candidate for further exploration in UAV networks \cite{info9070149}. However, these approaches do not fully explore the potential of advanced AI techniques in enhancing IDS capabilities for UAV networks.

\section{Proposed Approach}\label{sec:3}
This section introduces a robust system architecture (see Section~\ref{sec:3.1}) designed to secure communications between UAVs and ground stations. The proposed system architecture leverages the AES encryption alongside ECC and CRYSTALS-Kyber for key encapsulation, addressing the need for quantum-resistant encryption in a 5G environment. The architecture is designed to ensure secure transmission of video, audio, and image data while incorporating an AI framework for intrusion detection, enhancing the overall security of the communication system.

Then, we describe the proposed Data Flow and Processing (see Section~\ref{sec:3.2}) and explain the proposed dataset Generation and IDS Implementation (see Section~\ref{sec:3.3}).

\subsection{Proposed System Architecture}\label{sec:3.1}

\begin{figure*}[!ht]
\centering
\includegraphics[width=1.0\textwidth]{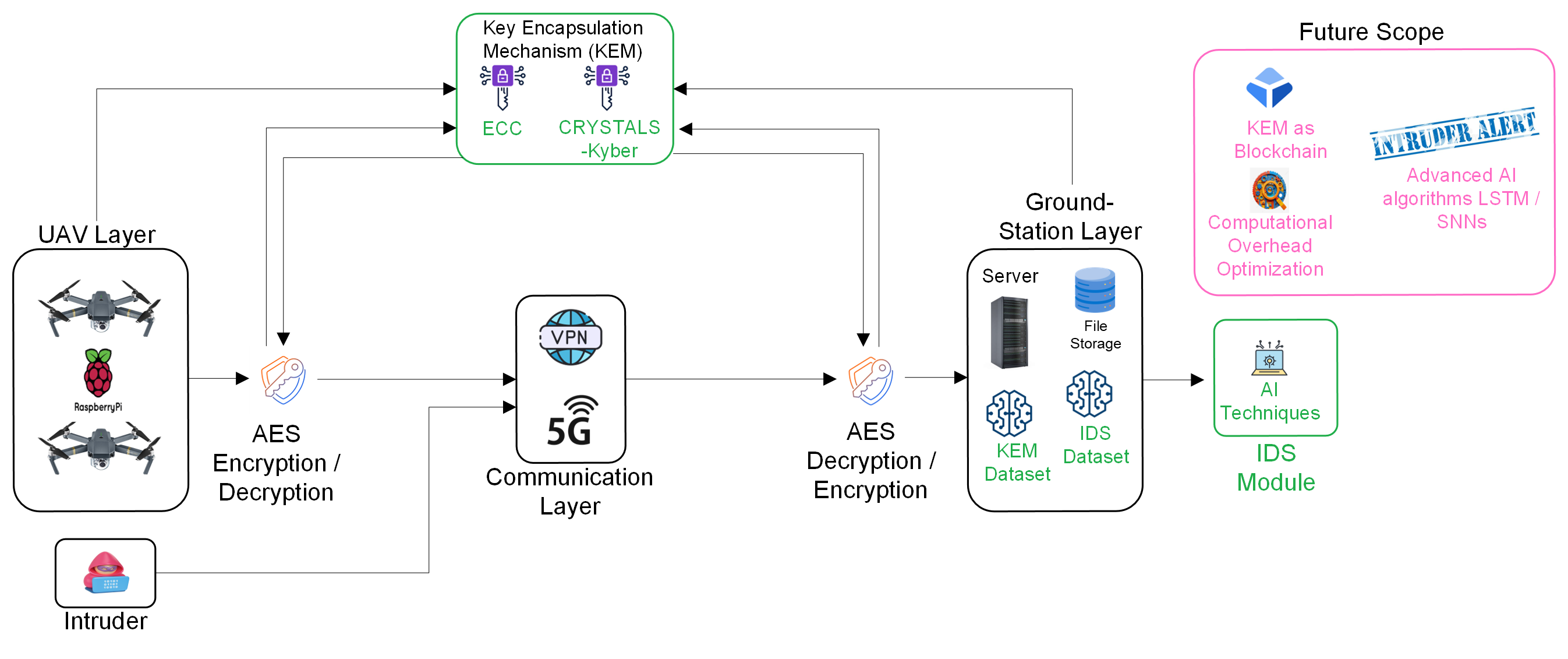}
\caption{\small The Proposed System Architecture Diagram}
\label{fig:fig1}
\end{figure*}

The proposed system architecture, displayed in Fig.~\ref{fig:fig1}, comprises a server communicating with multiple UAVs (clients) in a 5G environment. The server and UAVs (clients) select ECC and CRYSTALS-Kyber as the key encapsulation method, while AES in EAX encrypts the data. This dual-layered security approach ensures the confidentiality and integrity of the transmitted data, even in the presence of potential quantum attacks.

The architecture is structured into multiple layers, each serving distinct roles in securing UAV communication. At the UAV Layer, UAVs act as clients, equipped with Raspberry Pi SBCs and AES encryption modules in EAX mode to handle data processing and secure transmission. The Communication Layer facilitates the data transfer between UAVs and the Ground Station Layer via a 5G network or VPN, ensuring safe and versatile connectivity.

At the Ground Station Layer, the server decrypts the incoming data and stores it for further analysis. This layer also hosts the AI-powered IDS, which continuously monitors communication patterns between UAVs and the server and identifies potential intrusions using advanced machine learning techniques.

For enhanced security, the system employs Key Encapsulation Mechanisms (KEM), specifically ECC and CRYSTALS-Kyber, to protect against quantum computing threats. The architecture also allows for future enhancements, including integrating blockchain-based key management, advanced neural networks, and computational overhead optimization, ensuring long-term adaptability and resilience against emerging threats.

\begin{figure}[!ht]
\centering
\includegraphics[width=1\columnwidth]{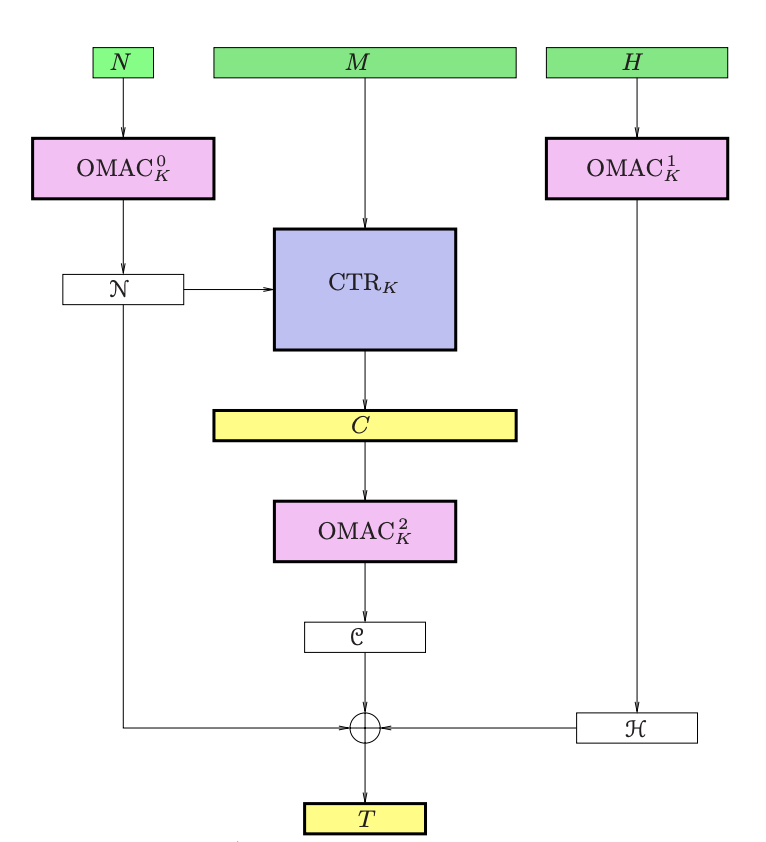}
\caption{Encryption under EAX \cite{Bellare2004}.}
\label{fig:fig3}
\end{figure}

Algorithms \ref{algorithms1}, \ref{algorithm2} specify the EAX algorithms used for encryption and decryption, while Figure \ref{fig:fig3} illustrates the EAX encryption process\cite{Bellare2004}.

\begin{algorithm}[t!]
\caption{EAX.Encrypt$^{N H}_{K} \hspace{0.1cm}(M)$}
\label{algorithms1}
\begin{algorithmic}[1]
    \State $\EuScript{N} \gets$ OMAC$_{K}^{0}(N)$
    \State $\EuScript{H} \gets$ OMAC$_{K}^{1}(H)$
    \State $C \gets$ CTR$_{K}^{\EuScript{N}}(M)$
    \State $\EuScript{C} \gets$ OMAC$_{K}^{2}(C)$
    \State $Tag \gets \EuScript{N} \oplus \EuScript{C} \oplus \EuScript{H}$
    \State $T \gets \text{Tag}[ \text{first } \tau \text{ bits}]$
    \State \textbf{return} $CT \gets C \parallel T$
\end{algorithmic} 
\end{algorithm}

\begin{algorithm}[t!]
\caption{EAX.Decrypt$^{N H}_{K} \hspace{0.1cm}(CT)$}
\label{algorithm2}
\begin{algorithmic}[1]
    \If{$|CT| < \tau$} \textbf{return} \textit{INVALID}
    \EndIf
    \State Let $C \parallel T \gets CT$ where $ |T| = \tau$
    \State $\EuScript{N} \gets$ OMAC$_{K}^{0}(N)$
    \State $\EuScript{H} \gets$ OMAC$_{K}^{1}(H)$
    \State $\EuScript{C} \gets$ OMAC$_{K}^{2}(C)$
    \State $Tag' \gets \EuScript{N} \oplus \EuScript{C} \oplus \EuScript{H}$
    \State $T' \gets Tag'[ \text{first } \tau \text{ bits}]$
    \If{$T \neq \text{T}'$} \textbf{return} \textit{INVALID}
    \EndIf
    \State $M \gets$ CTR$_{K}^{\EuScript{N}}(C)$
    \State \textbf{return} $M$
\end{algorithmic}

\end{algorithm}

``The plaintext is \(M\), the ciphertext is \(CT\), the key is \(K\), the nonce is \(N\), and the header is \(H\). The mode depends on a block cipher \(E\) (that CTR and OMAC implicitly use) and a tag length $\tau$'' \cite{Bellare2004}. OMAC is secure for messages of any bit length and provides an extra layer of security over CBC MAC - only secure on messages of fixed length, which is in multiple of the block length.

\noindent The message is \(M\), the key is \(K\), and the header is \(H\) \cite{Bellare2004}. The ciphertext is 
\begin{equation}
CT = C \parallel T
\end{equation}

\subsection{Proposed Data Flow and Processing}\label{sec:3.2}

The communication begins with establishing a secure connection between the UAVs and the ground station. Once the connection is established, the selected KEM (ECC or CRYSTALS-Kyber) is used to securto exchange keys between the clients and the server securely keys are then used to encrypt the data using AES before transmission. Upon receiving the data, the server decrypts it and stores it securely. 

\begin{figure*}[!ht]
\centering
\includegraphics[width=1.0\textwidth]{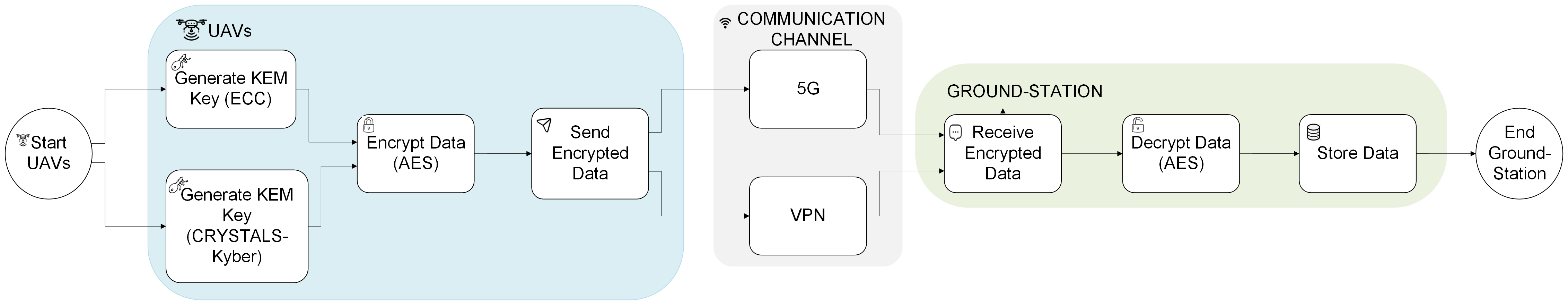}
\caption{\small The proposed data encryption algorithm.}
\label{fig:fig4}
\end{figure*}

Fig. ~\ref{fig:fig4} illustrates the data flow algorithm using KEMs, ECC and CRYSTALS-Kyber and AES, to encrypt the data between UAVs and Ground-Station.

To enhance security further, an IDS is integrated within the architecture. The IDS leverages AI techniques to analyze communication patterns and detect potential intrusions. Data from the communication process is logged and used to train machine learning models that can accurately identify and mitigate threats. This AI-driven approach to IDS has shown significant promise in both simulated and real-world environments, particularly in scenarios involving complex network topologies \cite{10418859, 10315971}.

\subsection{Proposed Dataset Generation and IDS Implementation}\label{sec:3.3}
The system records several parameters such as encryption time, decryption time, and server response time, which are used to create datasets for evaluating the performance of ECC and CRYSTALS-Kyber. Additionally, a dataset is generated to simulate real-world intrusion scenarios by introducing anomalies during the communication process. This dataset trains the IDS, enabling it to detect unauthorized access with high accuracy. Various studies have shown that using synthetic datasets for IDS training can lead to highly effective models, particularly in environments where real-time data is scarce or difficult to obtain.

The proposed approach combines quantum-resistant encryption with AI-based intrusion detection to create a comprehensive security framework for UAV communications in a 5G environment. The architecture is designed to be lightweight, making it suitable for deployment on SBCs like Raspberry Pi, commonly used in UAV systems. This design is consistent with recent trends in cybersecurity research, which emphasize the importance of scalability and efficiency in cryptographic solutions.

\section{Performance Evaluation}\label{sec:4}

This section presents the results of our performance evaluation, focusing on both the cryptographic mechanisms used (ECC and CRYSTALS-Kyber) and the effectiveness of the IDS across multiple experimental setups.

\subsection{Dataset}\label{sec:4.1}

Separate sets of datasets were recorded for the performance evaluation of the KEM and the IDS. These datasets provided a comprehensive basis for analyzing the security and efficiency of the proposed system architecture in VPN and 5G environments, illustrated in \ref{sec:4.1.1} and \ref{sec:4.1.2}.

\subsubsection{KEM Performance Dataset}\label{sec:4.1.1}

The KEM performance evaluation was conducted using two distinct datasets recorded through the application in the VPN and 5G environment \cite{sourceCode}. These datasets captured various performance metrics, including handshake time, encryption time, decryption time, connection duration, server response time, CPU usage, and memory usage. The experiments compared the performance of ECC and CRYSTALS-Kyber as KEMs in both VPN and 5G network conditions.

The same audio and video files were utilized in both network environments to ensure consistency across experiments. The results from these datasets highlight the trade-offs between performance and quantum-resistant security. Specifically, while CRYSTALS-Kyber exhibited a slight overhead in CPU usage and handshake time compared to ECC, this overhead was not significant. CRYSTALS-Kyber still demonstrates its viability as a lightweight KEM \cite{HE2024167}, making it a strong candidate for deployment in resource-constrained environments like UAVs, where quantum-resistant encryption is increasingly essential.

\subsubsection{IDS Performance Dataset}\label{sec:4.1.2}

For the IDS experiments, several datasets were generated to simulate communication between UAVs (clients) and a ground station (server), including normal operations and intrusion scenarios. The IDS datasets included records of communication behaviors, such as message size, file size, connection duration, and data transfer volume. The datasets were split into training and testing sets with varying ratios of clients to intruders to evaluate the IDS's ability to detect anomalies.

Key details of the IDS datasets are as follows:

\begin{itemize}[leftmargin=*]
    \item \textbf{Experiment\hspace{0.05cm}1:} The training dataset contained approximately 27,000 records, with a 60-40 client-to-intruder ratio. The test dataset used a 70-30 client-to-intruder ratio, both sampled per second, to maintain temporal accuracy.
    \item \textbf{Experiment\hspace{0.05cm}2:} The same training dataset was used, but the test dataset had a more challenging 93-7 client-to-intruder ratio, sampled per second.
    \item \textbf{Experiment\hspace{0.05cm}3:} The training dataset size increased to 200,000 records, with a 70-30 client-to-intruder ratio. No per-second sampling was applied, allowing a more extensive evaluation of the IDS's performance.
    \item \textbf{Experiment\hspace{0.05cm}4:} Similar to Experiment 3, but the test dataset had a 93-7 client-to-intruder ratio, testing the IDS's ability to handle highly imbalanced datasets.
\end{itemize}

The IDS datasets facilitated the evaluation of various AI models, including Support Vector Machines (SVM), LightGBM, Random Forest (RF), Naive Bayes (NB), Neural Networks (NN), Logistic Regression (LR), and XGBoost, across different experimental setups. These models were assessed based on metrics such as accuracy and AUC to determine their effectiveness in detecting intrusions in UAV communications.

Overall, these datasets provided a robust foundation for analyzing the security and efficiency of the proposed UAV communication framework, offering insights into the performance of quantum-resistant encryption methods and AI-powered IDS.

\subsection{Simulation Setup}\label{sec:4.2}
The experiments were conducted in Virtual Private Network (VPN) and 5G environments. The UAVs were simulated as clients, and the ground station acted as the server. Both ECC and CRYSTALS-Kyber were used as KEMs in different experiments to compare their performance. The system architecture also included an IDS that employed various AI techniques to detect potential intrusions. We utilized the Kyber implementation from the repository provided by Giacomo Pope \cite{pope2023kyber}, along with the pycryptodome and cryptography libraries \cite{sourceCode}.

We measured several parameters for each environment, including handshake time, encryption time, decryption time, connection duration, server response time, CPU usage, and memory usage. These parameters were then compared across different experimental setups. The program source code of the proposed algorithms is available in \cite{sourceCode}.

\subsection{Comparative Performance of ECC and CRYSTALS-Kyber}\label{sec:4.3}

This subsection compares the performance of ECC and CRYSTALS-Kyber in both Virtual Private Network (VPN) and 5G environments. The evaluation metrics focus on handshake time, encryption and decryption times, CPU and memory usage. These results provide insights into the trade-offs between security and performance when using quantum-resistant encryption methods in UAV communication systems.

\subsubsection{Experiment 1}\label{sec4:3.1}

In this experiment, we assessed the performance of ECC and CRYSTALS-Kyber within a VPN environment. The results indicate that CRYSTALS-Kyber has slightly higher handshake times compared to ECC. However, both mechanisms' encryption and decryption times are comparable, as illustrated in Table \ref{tab:tab1}.

The memory and CPU usage metrics provide further insights into the lightweight nature of these cryptographic systems. CRYSTALS-Kyber exhibits a marginally higher CPU usage than ECC, while the memory usage is slightly lower for CRYSTALS-Kyber in the VPN environment Table \ref{tab:tab1}.

Overall, these results suggest that both ECC and CRYSTALS-Kyber are viable options for securing communications in resource-constrained environments like UAVs, with CRYSTALS-Kyber offering the added benefit of quantum resistance.

\subsubsection{Experiment 2}\label{sec4:3.2}

In this experiment, conducted in a 5G environment, the results are consistent with those observed in the VPN setup. The handshake times for CRYSTALS-Kyber remain slightly higher, while the Encryption and decryption times show minimal differences between the two mechanisms \ref{tab:tab1}. The CPU and memory usage metrics in the 5G environment also align with the VPN results, reaffirming that CRYSTALS-Kyber can be deployed effectively in real-time communication systems like 5G networks.

Table \ref{tab:tab1} comprehensively compares the performance metrics for both ECC and CRYSTALS-Kyber across VPN and 5G environments.

\begin{table}[ht]
\caption{\small ECC and CRYSTALS-Kyber comparisons for VPN and 5G.\vspace{-10px}}
\begin{center}
\resizebox{\columnwidth}{!}{%
\begin{tabular}{|>{\centering\arraybackslash}m{4cm}|c|c|c|}
\hline
\textbf{Parameter} & \textbf{Environment} & \textbf{ECC} & \textbf{CRYSTALS-Kyber} \\\hline
\multirow{2}{*}{\centering \textbf{Handshake Time (ms)}} & VPN & $5.2$ & $10.71$ \\\cline{2-4}
& 5G & $5.31$ & $10.55$ \\\hline
\multirow{2}{*}{\centering \textbf{Encryption Time (s)}} & VPN & $82.162$ & $81.281$ \\\cline{2-4}
& 5G & $108.445$ & $110.118$ \\\hline
\multirow{2}{*}{\centering \textbf{Decryption Time (s)}} & VPN & $82.175$ & $81.292$ \\\cline{2-4}
& 5G & $108.455$ & $110.129$ \\\hline
\multirow{2}{*}{\centering \textbf{Connection Duration (s)}} & VPN & $430$ & $350$ \\\cline{2-4}
& 5G & $417$ & $460$ \\\hline
\multirow{2}{*}{\centering \textbf{Server Response Time (s)}} & VPN & $82.188$ & $81.313$ \\\cline{2-4}
& 5G & $108.475$ & $110.150$ \\\hline
\multirow{2}{*}{\centering \textbf{CPU Usage (\%)}} & VPN & $0.04$ & $0.06$ \\\cline{2-4}
& 5G & $0.04$ & $0.08$ \\\hline
\multirow{2}{*}{\centering \textbf{Memory Usage (MB)}} & VPN & $24.358$ & $30.914$\\\cline{2-4}
& 5G & $24.449$ & $18.324$ \\\hline
\end{tabular}
}
\label{tab:tab1}
\end{center}
\end{table}

\subsubsection{Summary for KEM Experiments}\label{sec4:3.3}

The comparative analysis of ECC and CRYSTALS-Kyber in both VPN and 5G environments demonstrates the feasibility of deploying quantum-resistant cryptography in UAV communication systems. CRYSTALS-Kyber, while showing slightly higher handshake times and CPU usage, provides enhanced security against quantum threats, making it a strong candidate for future-proof cryptographic solutions. Both KEMs exhibited similar performance in terms of encryption and decryption times, with minimal differences in memory usage. These findings suggest that CRYSTALS-Kyber and ECC can be effectively utilized in resource-constrained environments like UAVs, ensuring secure and efficient communication without significant trade-offs in performance.

\subsection{IDS Performance}\label{sec:4.4}

This subsection evaluates the performance of different AI models in detecting intrusions across multiple experimental setups. The experiments were designed to assess the IDS under varying ratios of client and intruder data, simulating different threat levels in the UAV network. The goal was to identify the most effective AI model for intrusion detection in a simulated environment, representing typical UAV communications.

\subsubsection{Experiment 1}\label{sec4:4.1}

In this experiment, the IDS was trained using a dataset containing approximately 27,000 records and evaluated on a test dataset with a 70-30 ratio of clients to intruders. The training and test datasets were sampled per second to maintain temporal accuracy. As displayed in figures \ref{fig:ids_acc_all} and \ref{fig:ids_auc_all}, among the various AI models evaluated, Logistic Regression emerged as the top performer, achieving a notable accuracy of 89.50\% and an Area Under the Curve (AUC) of 0.97, indicating its strong capability in distinguishing between clients and intruders. This demonstrates the robustness of Logistic Regression in scenarios with a balanced distribution of clients and intruders.

\subsubsection{Experiment 2}\label{sec4:4.2}

This experiment introduced a more challenging scenario, utilizing the same training dataset as in Experiment 1, but evaluating the IDS on a test dataset with a skewed 93-7 ratio of clients to intruders. Similar to Experiment 1, the test dataset was sampled per second to ensure consistency in the temporal data representation. Despite the increased difficulty, the IDS models demonstrated strong performance, particularly Logistic Regression and Neural Networks, which achieved AUC values of 0.97 and 0.86, respectively, as outlined in figures \ref{fig:ids_acc_all} and \ref{fig:ids_auc_all}. These results underscore the robustness of these models in handling imbalanced datasets.

\subsubsection{Experiment 3}\label{sec4:4.3}

In this experiment, the size of the training dataset was increased to 200,000 records without per-second sampling, maintaining a 70-30 client-intruder ratio. Under this larger dataset, the XGBoost and LightGBM models demonstrated the highest accuracy of 87.82\% and 87.85\% with an AUC of 0.68 and 0.74, respectively, outperforming other models, as seen in figures \ref{fig:ids_acc_all} and \ref{fig:ids_auc_all}. This experiment indicates the effectiveness of XGBoost in scenarios with larger datasets, where it can leverage its boosting mechanism to improve performance. Although LightGBM showed good results, its performance was inadequate in the first two experiments.

\subsubsection{Experiment 4}\label{sec4:4.4}

The final experiment modified the dataset to include a 93-7 client-intruder ratio with 200,000 records. XGBoost once again showeagain showed superior performance, achieving 97.33\% AUC of 0.94. This experiment highlights the model's ability to detect intrusions even in highly imbalanced datasets, making it a robust.It is real-world applications where intrusions are rare but critical to identify.

\begin{figure}[!t]
    \begin{subfigure}[b]{0.48\textwidth}
        \centering
        \includegraphics[width=\textwidth]{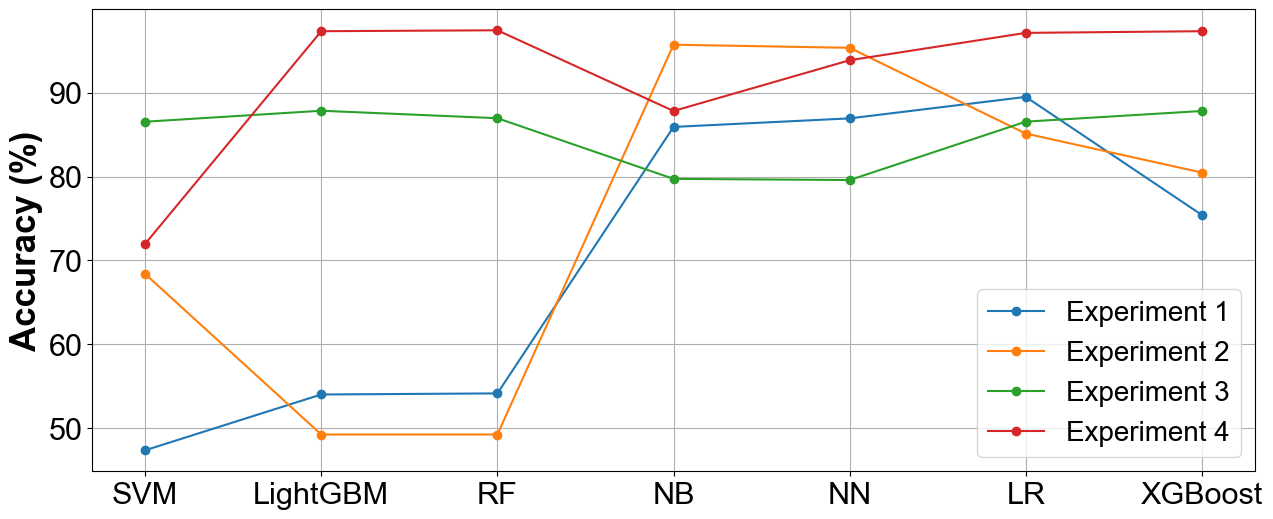}
        \caption{\small Accuracy performance.}
        \label{fig:ids_acc_all}
    \end{subfigure}
    \hfill
    \begin{subfigure}[b]{0.48\textwidth}
        \centering
        \includegraphics[width=\textwidth]{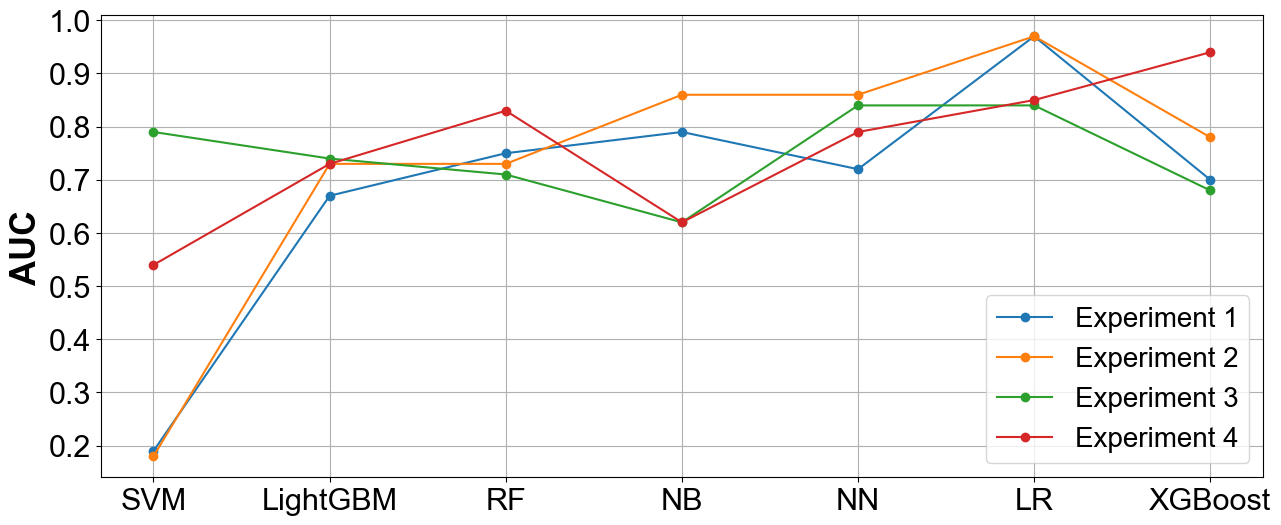}
        \caption{\small AUC performance.}
        \label{fig:ids_auc_all}
    \end{subfigure}
    \caption{\small Accuracy and AUC performances for different experiments.\vspace{-15px}}
\end{figure}

\noindent\textbf{Discussion for experiments:}\label{sec4:4.5}
The corresponding graphical representation of IDS accuracy and AUC across different models and experiments can be found in Figures \ref{fig:ids_acc_all} and \ref{fig:ids_auc_all}.
The graphs indicate that Logistic Regression and XGBoost consistently performed well across various experiments, with XGBoost showing solid performance in scenarios with larger datasets and more imbalanced data distributions. This suggests that while Logistic Regression is highly effective in more balanced scenarios, XGBoost may be better suited for environments with more complex and varied data distributions.
Integrating AI models within the IDS significantly improved the security of UAV communication systems. Among the evaluated models, XGBoost consistently delivered the best intrusion detection performance, especially in scenarios with highly imbalanced datasets, underscoring AI's potential to fortify UAV networks against diverse security threats.

\vspace{-4px}
\section{Conclusions and Future Works}\label{sec:5}

This paper presents a secure and efficient communication framework for UAV-to-ground station interactions, achieved through the integration of advanced cryptographic techniques and AI-driven IDS. A comprehensive comparison between ECC and CRYSTALS-Kyber for KEMs highlights their performance across various network environments, including VPN and 5G. While CRYSTALS-Kyber exhibits marginally higher resource consumption, it provides enhanced resilience against quantum attacks, establishing it as a future-proof cryptographic solution. Resource usage metrics further demonstrate the feasibility of implementing these encryption schemes on resource-constrained UAV systems, such as those utilizing SBCs. In future work, we will explore blockchain-based key encapsulation and encryption mechanisms, leveraging the decentralized and tamper-resistant nature of blockchain technology to enhance key management systems and mitigate risks of unauthorized access. Additionally, we will investigate the use of advanced neural network architectures, including Long Short-Term Memory (LSTM) networks and Spiking Neural Networks (SNNs), to more effectively capture temporal patterns and improve energy efficiency in IDS.

\section*{Acknowledgment}
This work is supported by Optimising mobile Network pErformance in High Demand Density Environments (ONE4HDD) Project, a major winners of the U.K's Department of Science, Innovation and Technology (DSIT) Open Networks Ecosystem Competition.

\bibliographystyle{IEEEtran}
\bibliography{refereces.bib}

\end{document}